\documentstyle[aps,epsf,twocolumn]{revtex}

\begin{document}
%\draft
%\tighten
\title
 {Telling Tails
 In The Presence
 Of A Cosmological Constant}
\author{
	Patrick R. Brady,${}^{(1)}$
        Chris M. Chambers,${}^{(2)}$
        William Krivan,${}^{(3,4)}$, and
        Pablo Laguna ${}^{(4)}$\\
       }
\address{
         ${}^{(1)}$ Theoretical Astrophysics,  Mail-Code 130-33,
         	  California Institute of Technology,
		  Pasadena,  CA 91125
        }
\address{
         ${}^{(2)}$ Department of Physics,
		   Montana State University,
        	   Bozeman,
		   MT 59717
        }
\address{
         ${}^{(3)}$ Institut f\"{u}r Astronomie und Astrophysik,
                   Universit\"{a}t T\"{u}bingen, D-72076 T\"{u}bingen,
                   Germany
        }
\address{
         ${}^{(4)}$ Department of Astronomy \& Astrophysics and
                   Center for Gravitational Physics \& Geometry,\\
                   Pennsylvania State University, University Park, PA 16802
        }
\date{\today}

%\preprint{GRP-XXX\\}
\maketitle

%%%%%%%%%%%%%%%%%%%%%%%%%%%%%%%%%%%%%%%%%%%%%%%%%%%%%%%%%%%%%%
%  ABSTRACT
%%%%%%%%%%%%%%%%%%%%%%%%%%%%%%%%%%%%%%%%%%%%%%%%%%%%%%%%%%%%%%

\begin{abstract}

We study the evolution of massless scalar waves propagating on
spherically symmetric spacetimes with a non-zero cosmological
constant.  Considering test fields on both Schwarzschild-de~Sitter
and Reissner-Nordstr\"{o}m-de~Sitter backgrounds, we demonstrate the
existence of {\em exponentially} decaying tails at late times.
Interestingly the $\ell=0$ mode asymptotes to a non-zero value,
contrasting the asymptotically flat situation.  We also compare these
results, for $\ell=0$, with a numerical integration of the
Einstein-scalar field equations, finding good agreement between the
two.  Finally, the significance of these results to the study of the
Cauchy horizon stability in black hole-de~Sitter spacetimes is
discussed.
\end{abstract}

%\tighten
\narrowtext

%%%%%%%%%%%%%%%%%%%%%%%%%%%%%%%%%%%%%%%%%%%%%%%%%%%%%%%%%%%%%%
%  SECTION 1  : Introduction
%%%%%%%%%%%%%%%%%%%%%%%%%%%%%%%%%%%%%%%%%%%%%%%%%%%%%%%%%%%%%%

\section{Introduction}

Perturbative studies of relativistic, spherical collapse have
elucidated dynamical features of gravitational collapse important to
an understanding of black hole formation and the subsequent relaxation
to a stationary state (see for example~\cite{mtw:73}).  Indeed,
quasi-normal ringing could provide direct evidence of the existence of
black holes if observed~\cite{kst:87}.  At late times, quasi-normal
oscillations are swamped by the radiative tail of the gravitational
collapse~\cite{price:72}.  This tail radiation is of interest in its
own right since it originates from the non-trivial propagation of the
field perturbations on the background spacetime---zero rest-mass
fields do not necessarily propagate along the light-cone in a
curved spacetime.

The first authoritative study of nearly spherical collapse, exhibiting
radiative tails, was performed by Price \cite{price:72}.  Studying the
behaviour of a massless scalar field propagating on a fixed
Schwarzschild background, he showed that the field dies off with the
now familiar power-law tail $t^{-(2\ell+P+1)}$ at late times, where
$\ell$ is the multipole order of the field, and $P=1$ if the field is
initially static and $P=2$ otherwise.  Furthermore, Price showed that
the perturbations of any zero rest-mass, integer-spin  field
are governed by a wave equation with the same qualitative form as that
governing the scalar field. This suggests
that the results for the scalar field  should apply equally well to the
radiatable multipoles of both the electromagnetic and gravitational fields.
Similar results for a massless scalar field propagating on a
Reissner-Nordstr\"{o}m background have been obtained by Bi\v{c}\'{a}k
\cite{bicak1:72}. No such analytic result has yet been obtained for
the case of a black hole with angular momentum, though
Krivan {\it et al.} \cite{krivan:96} have recently performed numerical work
which
suggests the power law tail holds independently of the angular
momentum of the black hole.

While test-field calculations are extremely compelling, it is
natural to ask to what extent linear analyses are representative of
dynamical gravitational collapse.  If either quasi-normal ringing, or
radiative tails, should be absent in non-linear collapse one might
view results of linear analyses with scepticism.  Advances in
numerical relativity make it possible to address this issue in the
spherically symmetric context.  G\'{o}mez and Winicour
\cite{gomez:92} studied the non-linear evolution of a
self-gravitating, spherically symmetric, massless scalar field
concluding that the scalar monopole moment decayed exponentially
rather than with the power-law predicted by the linear analyses.  More
recently, Gundlach, Price and Pullin \cite{gundlach1:94,gundlach2:94}
re-examined this problem. They were able to show that the frequencies
of quasi-normal oscillations, and decay rates of the power-law tails,
found in the numerical solutions, are in good agreement with the
predictions of perturbation theory (though one must go to sufficiently
late times in order to see tail effects -- this would, in part,
explain the null result of G\'{o}mez and Winicour).

The presence, and slow decay, of wave tails at late times is a key
ingredient leading to the instability of Cauchy horizons inside
charged and rotating black holes.  For black holes in asymptotically
flat spacetime, the inverse power-law decay of perturbing fields at
the event horizon has been used to provide initial data in
linear~\cite{chandra:82} and
non-linear~\cite{israel:90,ori:91,ori:92,pat2:95} studies of the black
hole interior.  In particular, the form of the wave tail is largely
responsible for the weakness of the mass-inflation
singularity inside charged black holes~\cite{ori:91}, and is believed
to have similar consequences inside rotating holes~\cite{ori:92}.

In contrast to the asymptotically flat case,
perturbative~\cite{mellor1:90,pat:92,chambers2:94} and
non-linear~\cite{sinha} studies indicate that the inner Cauchy horizon
of charged and rotating black holes immersed in asymptotically
de~Sitter space can be stable. That stability persists for a finite
volume of the parameter space suggests that these spacetimes violate
the letter (if not the spirit) of the cosmic censorship
hypothesis~\cite{penrose:68}.  The nature of the radiative tail of
perturbations at late times plays a major role in these analyses.
Plausible arguments suggest that an exponential decay of the
tails replaces the power law behaviour observed in asymptotically flat
space~\cite{pat:92,chambers2:94}, however, no detailed analysis of the
evolution of wave tails in asymptotically de~Sitter-black hole
spacetimes exists.  The present work reports on such a study for
non-rotating black holes.  Our primary motivation has been to obtain
correct boundary conditions on the radiation at the event horizon for
use in a numerical study of the internal structure of charged black
holes in de~Sitter space~\cite{chambers2:94}.

The paper is organized as follows: In section II we consider the
propagation of massless, minimally coupled scalar fields on the
Schwarzschild-de~Sitter and Reissner-Nordstr\"{o}m-de~Sitter
black-hole backgrounds. We derive the equation governing the scalar
test-field, and numerically integrate it.  Two independent numerical
codes were used throughout the linear analysis;  a null evolution
scheme following that of Gundlach~{\it et al.}~\cite{gundlach1:94},
and a Cauchy evolution scheme similar to that described by Krivan
{\it et al.} \cite{krivan:96}.   We found complete agreement between them.
Our results show that, except for the $\ell=0$
mode, the field falls off exponentially at the
cosmological and black-hole event horizons, and at future timelike
infinity.  The rate of decay depends upon the surface gravity,
$\kappa_1$, of the cosmological horizon [see~Eq.~(\ref{2.5})], and the
multipole order of the field.  In particular, as $t\rightarrow\infty$
\begin{equation}
\phi \sim e^{-\ell \kappa_1 t} \qquad , \qquad \ell>0
\; ,
\end{equation}
where $t$ is defined by Eq.~(\ref{2.1}).  For $\ell=0$, the field
approaches a constant value at late times rather than decaying.  A
suggestion of this behaviour can be found in the analysis of Chambers
and Moss~\cite{chambers2:94} and, as argued there, is similar to the
situation within the black hole interior.  In section III, we study
the non-linear evolution of a spherically symmetric, self-gravitating
scalar field by numerically integrating the coupled Einstein-scalar
field equations. Confining attention to spherical symmetry implies we
gain information solely about the $l=0$ mode of the field.  We find,
in accord with our linear analysis, that the field approaches a
constant value at the cosmological event
horizon, the black hole event horizon and future timelike infinity,
demonstrating that the results of our linear analysis are
indicative of the full theory. Further to this, we inspect the
behaviour of the field's stress-energy, showing that
   \begin{equation}
	\phi_{,r} \sim e^{-2 k u} \; ,
   \end{equation}
with $k\simeq\kappa_1$ to within about $12\%$.
In section IV we make some final comments
about the implications of our results for Cauchy horizon stability in
black-hole-de~Sitter spacetimes and the related issue of cosmic
censorship.

%%%%%%%%%%%%%%%%%%%%%%%%%%%%%%%%%%%%%%%%%%%%%%%%%%%%%%%%%%%%%
%  SECTION 2 : Linear perturbation analysis
%%%%%%%%%%%%%%%%%%%%%%%%%%%%%%%%%%%%%%%%%%%%%%%%%%%%%%%%%%%%%

\section{A Linear Analysis}
In this section, we consider a massless, minimally coupled, scalar
field propagating on a fixed Reissner-Nordstr\"{o}m-de~Sitter
background.  Since our considerations are limited to the black hole
exterior, it is clear that the Schwarzschild-de~Sitter case can always
be obtained by setting the charge $q=0$ (when this fails, we explicitly
include the corresponding formulae).

\subsection{The equations}
\begin{figure}
\leavevmode
\begin{center}
\begin{minipage}{0.5\textwidth}
\vspace*{-2cm}
\epsfxsize=0.6\textwidth \epsfysize=0.6\textwidth {\epsffile{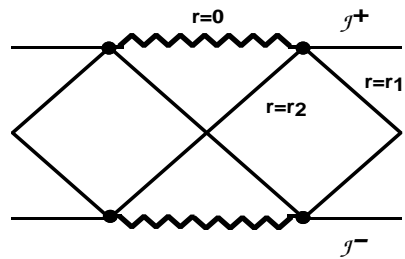}}
\end{minipage}
\end{center}
\vspace{0.5cm}
\caption{A conformal diagram representing the
Schwarzschild-de~Sitter  black hole spacetime. Shown are the
cosmological horizon at $r=r_{1}$, the black hole event horizon
at $r=r_{2}$ and  the singularity
(wavy line) located at $r=0$. Also shown are the locations
of past and future null infinity ${\cal J}^{-}$ and ${\cal J}^{+}$
respectively.}
\label{fig:1}
\end{figure}

\begin{figure}
\leavevmode
\begin{center}
\begin{minipage}{0.5\textwidth}
\vspace*{-2cm}
\epsfxsize=0.6\textwidth \epsfysize=0.6\textwidth {\epsffile{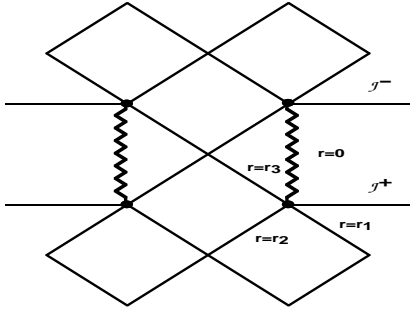}}
\end{minipage}
\end{center}
\vspace{0.5cm}
\caption{A conformal diagram showing the
Reissner-Nordstr\"{o}m-de~Sitter black hole spacetime. Shown are the
cosmological horizon at $r=r_{1}$, the black hole event horizon
at $r=r_{2}$, the inner (Cauchy) horizon at $r=r_{3}$
and  the singularity
(wavy line) located at $r=0$. Past and future null infinity
are indicated by ${\cal J}^{-}$ and ${\cal J}^{+}$ respectively.}
\label{fig:2}
\end{figure}

The generalization of the Reissner-Nordstr\"{o}m metric to include a
cosmological constant was first given by Carter~\cite{carter1:73} as
\begin{equation}
    ds^2=-f(r)dt^2+f(r)^{-1}dr^2+r^2(d\theta^2+\sin^{2}\theta
    d\phi^2) \; , \label{2.1}
\end{equation}
where
\begin{equation}
    f(r)=1-\frac{2M}{r}+\frac{q^2}{r^2}-\frac{r^2}{\alpha^2} \ \ \ , \ \ \
    \alpha^2=\frac{3}{\Lambda} > 0 \; . \label{2.2}
\end{equation}
In Eq.(\ref{2.2}), $M$ denotes the mass of the black hole, $q$ its charge and
$\Lambda$ is the cosmological constant.  If $q \neq 0$ there are three
horizons located at the roots of $f(r)=0$; an inner (Cauchy) horizon
at $r=r_{3}$, a black hole event horizon at $r=r_{2}$ and a
cosmological horizon located at $r=r_{1}$ such that
$r_{1}>r_{2}>r_{3}$ (see Fig.~\ref{fig:2}).  The fourth root,
$r_{4}$, is negative and thus non-physical.  (When $q=0$ there are only two
horizons (see Fig.~\ref{fig:1}) the black hole event horizon,
$r_{2}$, and the cosmological event horizon, $r_{1}$, while the third
root, $r_{3}$ is negative.)
%%%%%%%%%%%%%%%%%%%%%%%%%%%%%%%%%%%%%%%%%%%%%%%%%%%%%%%%%%%%%%%%%
%  FIGURE 1
%%%%%%%%%%%%%%%%%%%%%%%%%%%%%%%%%%%%%%%%%%%%%%%%%%%%%%%%%%%%%%%%%
%  FIGURE 2
%%%%%%%%%%%%%%%%%%%%%%%%%%%%%%%%%%%%%%%%%%%%%%%%%%%%%%%%%%%%%%%%%
%%%%%%%%%%%%%%%%%%%%%%%%%%%%%%%%%%%%%%%%%%%%%%%%%%%%%%%%%%%%%%%%%%%%%%%
\newpage
\widetext
It is convenient to introduce a ``tortoise'' radial coordinate,
$r_{*}=\int\! {dr}/{f(r)}$, which takes the explicit form
\begin{eqnarray}
           r_{*}&=&\left\{ \begin{array}{lr} {\displaystyle
	   -\frac{1}{2\kappa_{1}}\ln \left| \frac{r}{r_{1}}-1
	   \right|+\frac{1}{2\kappa_{2}}\ln \left| \frac{r}{r_{2}}-1
	   \right|-\frac{1}{2\kappa_{3}}\ln \left| \frac{r}{r_{3}}-1
	   \right|+\frac{1}{2\kappa_{4}}\ln \left| \frac{r}{r_{4}}-1
           \right|}  &  , \ \ \ q\neq 0 \: , \\ \\ {\displaystyle
	   -\frac{1}{2\kappa_{1}}\ln \left| \frac{r}{r_{1}}-1
	   \right|+\frac{1}{2\kappa_{2}}\ln \left| \frac{r}{r_{2}}-1
	   \right|+\frac{1}{2\kappa_{3}}\ln \left| \frac{r}{r_{3}}-1
	   \right|}   & , \ \ \ q=0 \; , \label{2.3}
	   \end{array} \right.
\end{eqnarray}
\narrowtext
where the arbitrary constant of integration has been set to zero.
We define
	\begin{equation}
	    \kappa_{i}=\frac{1}{2}\left| \frac{df(r)}{dr}\right|_{r=r_{i}}
	    \; , \label{6}
	\end{equation}
where $r_i$ are the roots of $f(r)=0$.  When the root corresponds to
a physical horizon in the spacetime, $\kappa_i$ is the surface gravity
of that horizon~\cite{wald:84}. Finally,   we introduce a pair of null
coordinates on the spacetimes,
the advanced time $v=t+r_{*}$ and the retarded time $u=t-r_{*}$, in
terms of which the interval (\ref{2.1}) reduces to,
\begin{equation}
    ds^2=-f(r) dudv
        +r^2(d\theta^2+\sin^{2}\theta d\phi^2) \; .
	  \label{2.4}
\end{equation}
The definition we have adopted means that the future cosmological horizon
$r=r_1$ is located at $v=\infty$,  and the future black-hole event
horizon $r=r_{2}$ is at $u=\infty$.  In terms of the null coordinates $(u,v)$,
the scalar wave equation, $\Box \phi=0$, becomes
\begin{equation}
    \Psi_{,uv} = -\frac{1}{4}V_{\ell}(r) \Psi \label{2.5a}
\end{equation}

\vspace*{3cm}
\noindent
\vrule height10pt width0.7pt \vskip-10pt \hrule height0.7pt width245pt

\begin{figure}
\leavevmode
\hbox{\epsfxsize=8cm \epsfysize=7cm {\epsffile{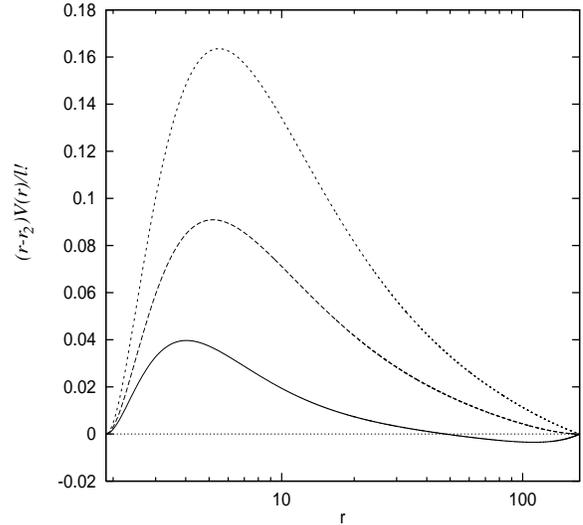}}}
\vspace{0.2cm}
\caption{The `effective' potential when $q=0.5$, $M=1$,
$\Lambda=10^{-4}$ and $\ell=0$ (solid), $\ell=1$ (dashed) and $\ell=2$
(dotted).  We have multiplied by $|r-r_2|/\ell!$ in order to
accentuate the nature of the potential when $\ell = 0$---the usual
potential barrier is followed by a potential well, a feature not
evident when $\ell>0$.  The potential tends to zero exponentially
quickly as $r_{*} \rightarrow \pm \infty$.}
\label{fig:3}
\end{figure}

\vspace{0.5cm}
\noindent
where we have decomposed the field $\phi$ into its constituent
multipole pieces, i.e. $\phi=\sum \Psi(u,v) Y_{\ell m}(\theta,\phi)
r^{-1}$.  The effective potential
\begin{equation}
    V_{l}(r)=f(r)
             \left(\frac{\ell(\ell+1)}{r^2}+\frac{2M}{r^3}
            -\frac{2q^{2}}{r^{4}}-\frac{2}{\alpha^2}\right) \; ,
             \label{2.5}
\end{equation}
is highly localized near to $r_*=0$,
falling off exponentially in $r_*$ at both $r=r_{1}$ and $r=r_{2}$.  The
form of the potential for $\ell=0,1$ and $\ell=2$ is shown in
Fig.~\ref{fig:3}.
%%%%%%%%%%%%%%%%%%%%%%%%%%%%%%%%%%%%%%%%%%%%%%%%%%%%%%%%%%%%%
%  FIGURE 3
%%%%%%%%%%%%%%%%%%%%%%%%%%%%%%%%%%%%%%%%%%%%%%%%%%%%%%%%%%%%%
%%%%%%%%%%%%%%%%%%%%%%%%%%%%%%%%%%%%%%%%%%%%%%%%%%%%%%%%%%%%%%%%%%%%%%%

%%%%%%%%%%%%%%%%%%%%%%%%%%%
% Linear Results
%%%%%%%%%%%%%%%%%%%%%%%%%%%

\subsection{The results}
It is straightforward to integrate Eq.(\ref{2.5a}) on a null grid using
the methods described by Gundlach {\it et al.} \cite{gundlach1:94}.
Further details can be found in their article.

Since tail effects are independent of the initial data used,  we
chose to represent a generic initial perturbation by a Gaussian pulse
on $u=0$
	\begin{equation}
	\Psi(u=0,v) = A \exp\{ -(v-v_1)^2/\sigma^2\} \; .
	\end{equation}
(The amplitude $A$ is irrelevant since Eq.~(\ref{2.5a}) is linear.
The data used to produce the figures had center $v_1=50.0$, and width
$\sigma=3.0$.)  The field is constant on $v=0$, $\Psi(v=0) =
\Psi(u=0,v=0)$.  We have set the mass of the hole, $M$, equal to unity
throughout; this corresponds to the freedom to rescale the coordinates
by an overall length scale. Investigations have shown that the results
are qualitatively similar for all $\Lambda > 0$, so we fix
$\Lambda=10^{-4}$ from here on.  We discuss the behaviour of the
field, $\phi=\Psi/r$, in three regions:
(a) timelike infinity---approached on surfaces of constant $r$.
(b) The cosmological horizon---in practice, approximated by the
null surface $v=v_{\mbox{\rm\scriptsize max}}$, the largest value
of $v$ on our grid.
(c) The black-hole event horizon---again,  approximated by the
null surface $u=u_{\mbox{\rm\scriptsize max}}$,  the largest
value of $u$ on our grid.

Gundlach {\it et al.}~\cite{gundlach1:94} argued that the nature of the
tails in asymptotically flat spacetimes is primarily due to the
power-law form of the effective potential as $r_*\rightarrow\infty$.
Moreover, Ching {\it et al.}~\cite{suen} have demonstrated that
inverse power-law tails,  as seen in spherically symmetric,
asymptotically flat spacetimes,  are not generic.  Therefore,  it
should not be surprising that tails, in asymptotically de~Sitter
spacetimes,  fall off exponentially with time since the effective
potential is exponentially suppressed as $r_*\rightarrow\pm\infty$.
What is unexpected is that the the $\ell=0$ modes do not decay to
zero,  rather a generic perturbation leads to a residual constant field
at late times.  (Of course,  there is no stress-energy associated
with a constant field.)  This was suggested in a paper by Chambers and
Moss~\cite{chambers2:94} where it was argued that the situation in
Schwarzschild-de~Sitter spacetime is somewhat analogous to scattering
inside a black hole~\cite{chandra:82},  and hence a constant mode can be
transmitted to both the black-hole and cosmological event horizons.
Indeed, the field at late times behaves like
	\begin{equation}
	\phi |_{l=0} \simeq \phi_0 + \phi_1(r) e^{-2\kappa_1 t}
		\; . \label{11}
	\end{equation}
%%%%%%%%%%%%%%%%%%%%%%%%%%%%%%%%%%%%%%%%%%%%%%%%%%%%%%%%%%%%%%%
%  FIGURE 4
%%%%%%%%%%%%%%%%%%%%%%%%%%%%%%%%%%%%%%%%%%%%%%%%%%%%%%%%%%%%%%%%%
\vspace*{-0.6cm}
\begin{figure}
\leavevmode
\noindent
%\vspace*{-0.2cm}
\hbox{\epsfxsize=8cm \epsfysize=3.3cm {\epsffile{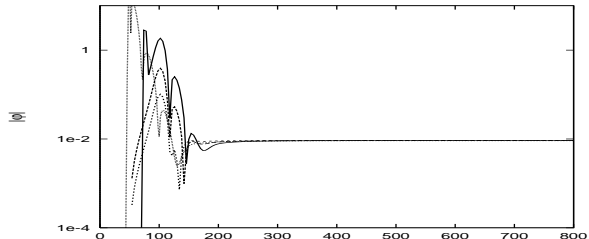}}}
\caption{A plot of $|\phi |$ versus time for
$q=0.5$, $\ell=0$, $M=1$ and $\Lambda=10^{-4}$.  The field is
shown on the cosmological event horizon $r_{1}$ (short-dashed), the
black hole event horizon $r_{2}$ (dotted) and two surfaces of constant
radius (solid and long-dashed).  Initially the quasi-normal
ringing dominates, while at late times the field settles down to the
same constant value on all four surfaces.}
\label{fig:4}
\end{figure}
Figure~\ref{fig:4} demonstrates this effect.
The early time behaviour of the field,
$\phi$, is dominated by the quasi-normal ringing, whose explicit features
we are not currently interested in. At later times the field
quickly approaches the same constant value at  the cosmological
event horizon, the black-hole event horizon and future timelike
infinity.
%%%%%%%%%%%%%%%%%%%%%%%%%%%%%%%%%%%%%%%%%%%%%%%%%%%%%%%%%%%%%%%%%%%%%%

\begin{figure}
\leavevmode
\hbox{\epsfxsize=8cm \epsfysize=7cm {\epsffile{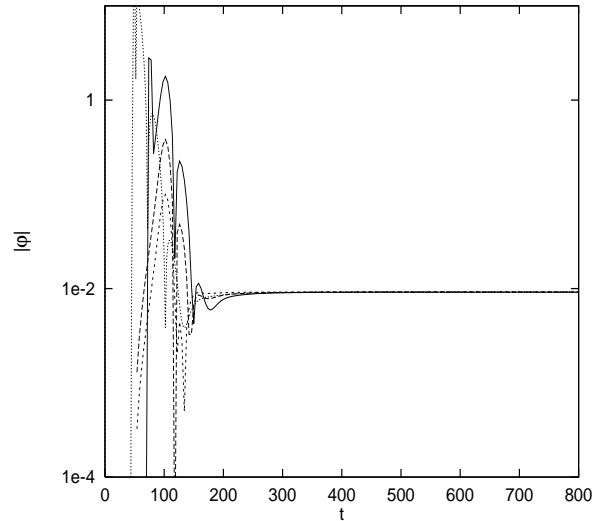}}}
\caption{A plot of $|\phi |$ versus time when
$q=0.0$, $\ell=0$, $M=1$ and $\Lambda=10^{-4}$.  The field is exhibited
on the same surfaces as the $q=0.5$ cases.  Once again the
asymptotically constant field is evident.}
\label{fig:7}
\end{figure}

One might suspect that the existence of a non-zero field at
late times is an artifact of the initial data,  or that setting
$q=0$ might produce different results.  Therefore Fig.~\ref{fig:7}
shows results for a Schwarzschild-de~Sitter  black hole with $\ell=0$.
Once again, the field settles down to a constant value which is
independent of the radial position.    One must contrast this to
previous results for asymptotically flat spacetimes where the field
always approaches zero.
Monitoring the behaviour of $\phi_{,t}$ during the evolution
(this is a matter of formality in the Cauchy code, as it uses
momenta $p_{t}= \phi_{,t}$ for the evolution) has allowed us to
conclude that the numerical
value of the exponent in Eq.(\ref{11}) is indeed $\kappa_{1}$ to
within $10 \%$: Taking the time derivative of  Eq.(\ref{11})
yields $\ln|\phi_{,t}|=\ln(2\kappa_1 \phi_1(r)) - 2 \kappa_1 t$, thus
the fitted values for $\kappa_1$ can be compared with the ones
obtained from Eq.(\ref{6}).

\begin{figure}
\leavevmode
\hbox{\epsfxsize=8cm \epsfysize=8cm {\epsffile{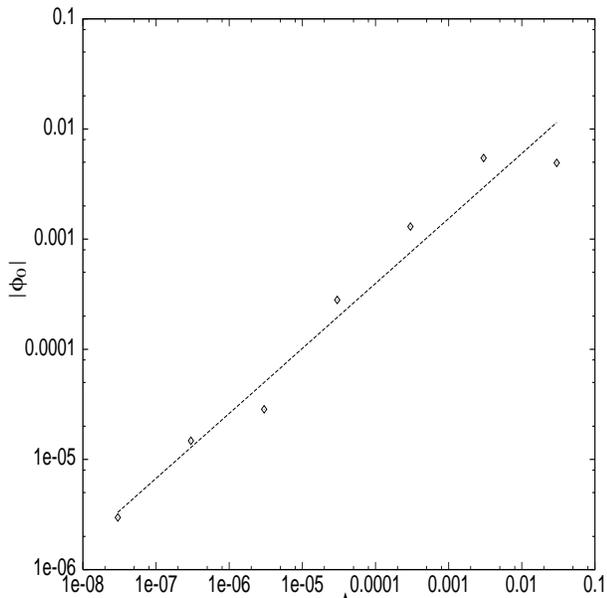}}}
\caption{A plot of $|\phi_0 |$ versus $\Lambda$
for $q=0.0$, $\ell=0$, $M=1$.
The linear correlation coefficient $c_{\mbox{\rm lin}}$
of $\ln |\phi_0|$ versus $\ln \Lambda$ is given by $c_{\mbox{\rm lin}} =
0.98$.
The dashed line represents the least square fit to $|\phi_0 | = b
\,\Lambda^a$  with $a=0.59$, $b=0.09$.}
\label{fig:lambdafit}
\end{figure}

In general, the final field value, $\phi_{0}$, is a function of the
black hole parameters $(M,q,\Lambda)$.
We investigated the dependance of $\phi_{0}$ on the cosmological
constant $\Lambda$.
As shown in Fig.~\ref{fig:lambdafit} for the case $q=0, M=1$,
our results indicate that $\phi_{0}\sim \Lambda^{0.6}$.
The linear correlation coefficient $c_{\mbox{\rm\scriptsize lin}}$
of $\ln |\phi_0|$ versus $\ln \Lambda$ is given by
$c_{\mbox{\rm\scriptsize lin}} =0.98$.

%%%%%%%%%%%%%%%%%%%%%%%%%%%%%%%%%%%%%%%%%%%%%%%%%%%%%%%%%%%%%%%
%  FIGURE 5
%%%%%%%%%%%%%%%%%%%%%%%%%%%%%%%%%%%%%%%%%%%%%%%%%%%%%%%%%%%%%%%%%
%%%%%%%%%%%%%%%%%%%%%%%%%%%%%%%%%%%%%%%%%%%%%%%%%%%%%%%%%%%%%%%%%%%%%%

%%%%%%%%%%%%%%%%%%%%%%%%%%%%%%%%%%%%%%%%%%%%%%%%%%%%%%%%%%%%%%%
%  FIGURE 6
%%%%%%%%%%%%%%%%%%%%%%%%%%%%%%%%%%%%%%%%%%%%%%%%%%%%%%%%%%%%%%%%%
%%%%%%%%%%%%%%%%%%%%%%%%%%%%%%%%%%%%%%%%%%%%%%%%%%%%%%%%%%%%%%%%%%%%%%

\begin{figure}
\leavevmode
\hbox{\epsfxsize=8cm \epsfysize=7cm 
{\epsffile{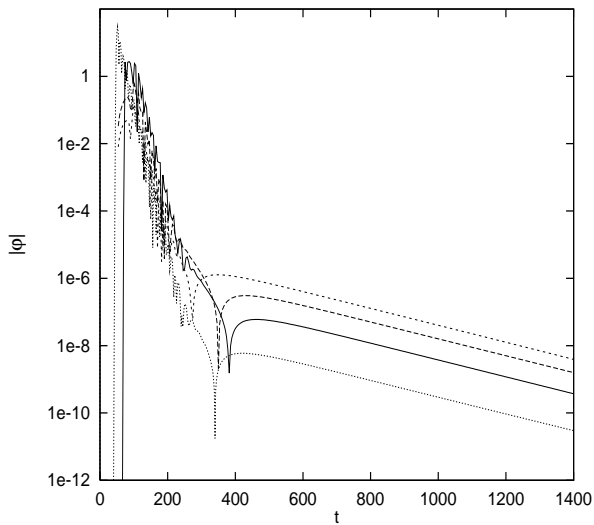}}}
\caption{A plot of $|\phi |$ versus time for
$q=0.5$, $\ell=1$, $M=1$ and $\Lambda=10^{-4}$.  The field along is
shown on the cosmological event horizon $r_{1}$ (short-dashed), the
black hole event horizon $r_{2}$ (dotted) and two surfaces of constant
radius (solid and long-dashed). The field falls off as $\exp{(-k t)}$
at late times,   with $k\simeq\kappa_1$ to within about $2\%$.   Note that
the ordinate scale is logarithmic.}
\label{fig:5}
\end{figure}

\begin{figure}
\leavevmode
\hbox{\epsfxsize=8cm \epsfysize=7cm {\epsffile{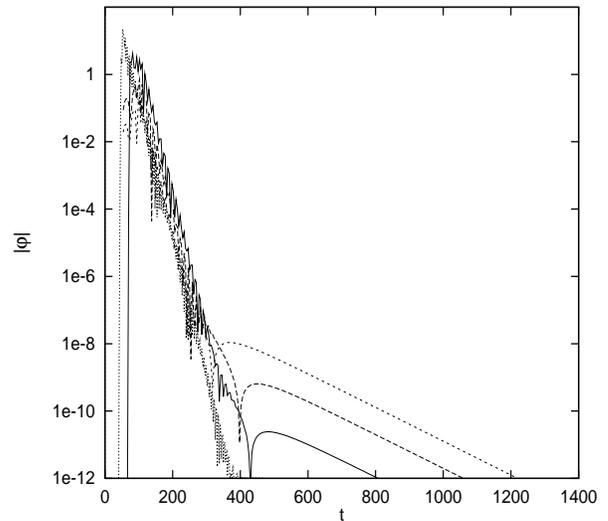}}}
\caption{A plot of $|\phi |$ versus time for
$q=0.5$, $\ell=2$, $M=1$ and $\Lambda=10^{-4}$.  The field is shown on
the cosmological event horizon $r_{1}$ (short-dashed), the black hole
event horizon $r_{2}$ (dotted) and two surfaces of constant radius
(solid and long-dashed).  At early times quasi-normal ringing
completely dominates, but eventually the field falls off as
$\exp{(-2k t)}$.  The value of $k\simeq\kappa_1$ is again accurate to
approximately $2\%$.  Note that the ordinate scale is logarithmic.}
\label{fig:6}
\end{figure}

For $\ell>0$ the picture is different, as seen from Fig.~\ref{fig:5}
and Fig.~\ref{fig:6}.  The early time behaviour of the field is still
dominated by complicated quasi-normal ringing, but at late times a
definite exponential fall-off is manifest. In particular, the late time
wave tails are well approximated by $\phi|_\ell \sim
\exp(-\ell\kappa_1 t)$. In a series of separate evolutions we
found that for sufficiently small values of $\Lambda$, a regime
of power law decay followed the quasi-normal ringing, though at
late times the exponential decay described above always dominates.

%%%%%%%%%%%%%%%%%%%%%%%%%%%%%%%%%%%%%%%%%%%%%%%%%%%%%%%%%%%%%%%%%
%  FIGURE 7
%%%%%%%%%%%%%%%%%%%%%%%%%%%%%%%%%%%%%%%%%%%%%%%%%%%%%%%%%%%%%%%%%
%  FIGURE 8
%%%%%%%%%%%%%%%%%%%%%%%%%%%%%%%%%%%%%%%%%%%%%%%%%%%%%%%%%%%%%%%%%%%
%%%%%%%%%%%%%%%%%%%%%%%%%%%%%%%%%%%%%%%%%%%%%%%%%%%%%%%%%%%%%%%%%%%%%%
%%%%%%%%%%%%%%%%%%%%%%%%%%%%%%%%%%%%%%%%%%%%%%%%%%%%%%%%%%%%%%%%%%%%%%

%%%%%%%%%%%%%%%%%%%%%%%%%%%%%%%%%%%%%%%%%%%%%%%%%%%%%%%%%%%%%
%  SECTION 3 : Non-Linear evolution
%%%%%%%%%%%%%%%%%%%%%%%%%%%%%%%%%%%%%%%%%%%%%%%%%%%%%%%%%%%%%

\section{Non-Linear Analysis}
Given the somewhat unusual behaviour of the $\ell=0$ modes elucidated by
the test field analysis,   it seems necessary  to examine the
non-linear evolution of a self-gravitating massless, scalar field
in the presence of a cosmological constant.   This situation is
described by the coupled Einstein-scalar field equations
\begin{equation}
    G_{\alpha \beta}=8 \pi T_{\alpha \beta}
	- g_{\alpha \beta} \Lambda \; , \label{3.1}
\end{equation}
where
\begin{equation}
    T_{\alpha \beta}=\phi_{,\alpha}\phi_{,\beta} -
    (1/2) g_{\alpha \beta} (\phi_{,\gamma} \phi^{,\gamma}) \; ,
    \label{3.3}
\end{equation}
is the stress-energy of the scalar field which satisfies $\Box\phi =0$.
Restricting attention to spherical symmetry (thus we only gain
information about the $\ell=0$ mode)  the line element can be written as
\begin{equation}
    ds^{2}= -g\bar{g} du^{2}-2gdudr+r^2(d\theta^{2}+\sin^{2}\theta
	    d\phi^{2}) \; , \label{3.4}
\end{equation}
where $g=g(u,r)$, $\bar{g}=\bar{g}(u,r)$ and $u$ is the retarded
time. The coordinates have been normalized so that $u$ is the
proper time at the origin, thus $g(0,r)=\bar{g}(0,r)=1$.
It is customary \cite{christo1:86} to introduce two new fields
$\bar{h}(u,r)$ and $h(u,r)$ defined by
\begin{equation}
    \phi=\bar{h}=\frac{1}{r}\int^{r}_{0} h \; dr \; . \label{3.5}
\end{equation}
In terms of these variables the field equations (\ref{3.1}) become
\begin{eqnarray}
      (\ln g)_{,r}&=&4\pi r^{-1}(h-\bar{h})^{2} \; , \label{3.6a}\\
      (r \bar{g})_{,r}&=&g(1-\Lambda r^2) \; , \label{3.6b}\\
      (r\bar{h})_{,r}&=& h \; , \label{3.6c}
\end{eqnarray}
and the wave equation is
\begin{equation}
     h_{,u}-\frac{\bar{g}}{2}h_{,r}=\frac{(h-\bar{h})}{2r}\left[
     g(1-\Lambda r^2)-\bar{g} \right] \; . \label{3.7}
\end{equation}
A well established numerical algorithm exists to integrate these
equations~\cite{gundlach2:94,goldwirth:87,garfy:95}.  We refer the
reader to these articles for details.

%%%%%%%%%%%%%%%%%%%%%%%%%%%%%%%%%%%%
% SUBSECTION : Non-Linear Results
%%%%%%%%%%%%%%%%%%%%%%%%%%%%%%%%%%%%

\subsection{The results}

Initial data for the non-linear equations is supplied on an
initial null cone centered on the origin.   We considered a
Gaussian pulse on $u=0$, with an amplitude $\phi_{0} (r/r_0)^2$,
width $\sigma$, and centered on $r=r_{0}$.  The code was tested
against the exact solution in~\cite{pat:94}.  We also reproduced
power-law tails in complete agreement with \cite{gundlach2:94},  when
$\Lambda=0$.

\begin{figure}
\leavevmode
\hbox{\epsfxsize=8cm \epsfysize=7cm {\epsffile{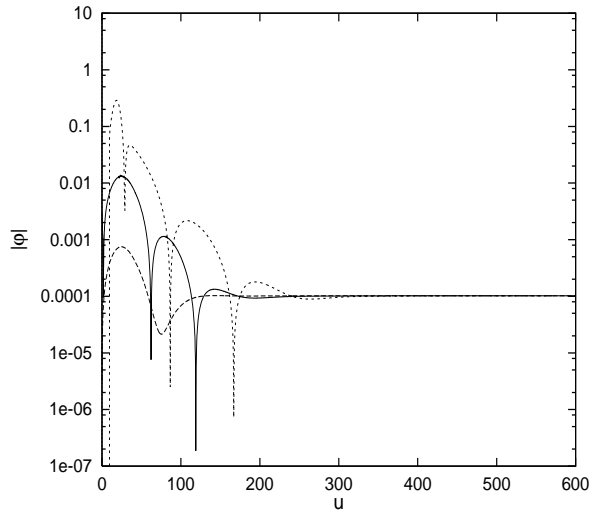}}}
\caption{A plot of $| \phi |$ versus time for
$\Lambda=10^{-4}$.  The field is shown along three surfaces: the
cosmological event horizon $r_1$ (short-dashed), the black hole event
horizon $r_{2}$ (long-dashed) and a surface of constant radius
(solid).  After a period of quasi-normal ringing the field settles
down to the same constant value at all three surfaces. }
\label{fig:8}
\end{figure}

Figure~\ref{fig:8} shows $\bar{h}$ (or equivalently $\phi$)
at the cosmological event horizon, the black hole event horizon
and along an $r=constant$ surface. The agreement with the linear
analysis is remarkable.   We again see an initial period of quasi-normal
ringing which decays,  leaving behind a constant field at late
times.
%%%%%%%%%%%%%%%%%%%%%%%%%%%%%%%%%%%%%%%%%%%%%%%%%%%%%%%%%%%%%%%%%%%
%  FIGURE 9
%%%%%%%%%%%%%%%%%%%%%%%%%%%%%%%%%%%%%%%%%%%%%%%%%%%%%%%%%%%%%%%%%%%%%
%%%%%%%%%%%%%%%%%%%%%%%%%%%%%%%%%%%%%%%%%%%%%%%%%%%%%%%%%%%%%%%%%%%%%%

\begin{figure}
\leavevmode
\hbox{\epsfxsize=8cm \epsfysize=7cm {\epsffile{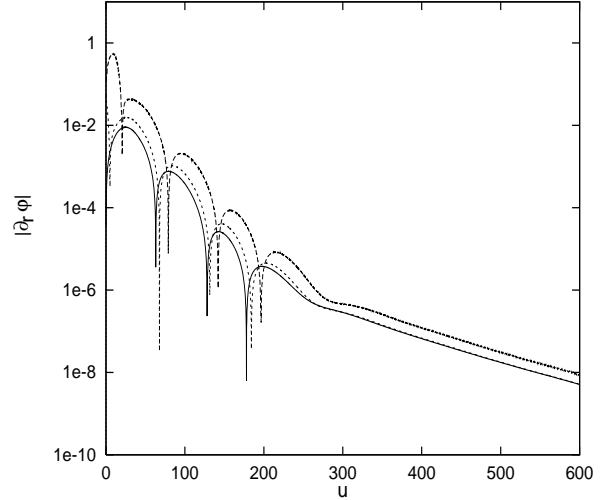}}}
\caption{A plot of $ |\phi_{,r}| $ versus time for
$\Lambda=10^{-4}$ along the same three surfaces.  At late times
exponential decay with the approximate form $\exp{(-2 k u)}$ is
evident---we have determined the value of $k\simeq\kappa_1$ within
about $12\%$. }
\label{fig:9}
\end{figure}

Furthermore,  using (\ref{3.5}) and (\ref{3.6c}) we can write
$r\phi_{,r}= (\bar{h}-h)$ and examine the stress-energy of the field
at late times.   Figure~\ref{fig:9}  suggests that $(\bar{h}-h) \sim
\exp[-2\kappa_1 u]$ at late times,  as suggested by the the test-field
analysis.
%%%%%%%%%%%%%%%%%%%%%%%%%%%%%%%%%%%%%%%%%%%%%%%%%%%%%%%%%%%%%%%%%%%%%
%  FIGURE 10
%%%%%%%%%%%%%%%%%%%%%%%%%%%%%%%%%%%%%%%%%%%%%%%%%%%%%%%%%%%%%%%%%%%
%%%%%%%%%%%%%%%%%%%%%%%%%%%%%%%%%%%%%%%%%%%%%%%%%%%%%%%%%%%%%
%  SECTION 5 :  Remarks
%%%%%%%%%%%%%%%%%%%%%%%%%%%%%%%%%%%%%%%%%%%%%%%%%%%%%%%%%%%%%

\section{Concluding Remarks}
This study of wave-tail evolution in black hole-de~Sitter spacetimes
has revealed consistent but interesting results.  The radiative tails
associated with a massless, minimally coupled scalar field propagating
on the fixed backgrounds of a Schwarzschild-de~Sitter and
Reissner-Nordstr\"{o}m-de~Sitter black hole decay exponentially at the
cosmological horizon, the black hole event horizon and at future
timelike infinity.  That is, for all modes except the $\ell=0$
mode, it approaches a constant at late times.  We have further
explored this result by considering the non-linear problem of a scalar
field coupled to gravity and numerically integrating the
Einstein-scalar field equations. These results show excellent
agreement with the linear analysis.

Though the introduction of a non-zero cosmological constant into the
Einstein equations may be argued to be somewhat unrealistic, it is the
issue of Cauchy horizon stability in black hole-de~Sitter
spacetimes~\cite{mellor1:90,pat:92,chambers2:94} and the related issue
of cosmic censorship~\cite{penrose:68}, that motivates this
investigation.  The power-law tails found by Price \cite{price:72}
have been used as initial data in arguments pertaining to the
instability of the inner (Cauchy) horizon of both the
Reissner-Nordstr\"{o}m and Kerr black hole spacetimes and the
associated phenomenon of mass inflation.  Linear perturbation studies
suggest that the Cauchy horizon inside black holes embedded in
de~Sitter space may be stable---fluxes of linear perturbations remain
bounded at the Cauchy horizon.  This in itself does not guarantee
stability though.  A fully non-linear analysis of the black hole
interior is needed, for which the late-time wave tails at the event
horizon serve as initial data.  With the results of this paper in
hand, we can now embark on a numerical study of the Cauchy horizon in
Reissner-Nordstr\"{o}m-de~Sitter, along the lines of Brady and
Smith~\cite{pat2:95}.

%%%%%%%%%%%%%%%%%%%%%%%%%%%%%%%%%%%%%%%%%%%%%%%%%%%%%%%%%%%%%
%  SECTION 6 : Acknowledgments
%%%%%%%%%%%%%%%%%%%%%%%%%%%%%%%%%%%%%%%%%%%%%%%%%%%%%%%%%%%%%

\acknowledgments

PRB is grateful to Bruce Allen and Kip Thorne for helpful discussions.
CMC wishes to thank the Caltech Relativity Group for hospitality during the
completion of this work and the Relativity group at Montana
State University for their constant support over the last year.
CMC is a Fellow of The Royal Commission For The Exhibition Of 1851
who's financial support is gratefully acknowledged.
PRB is supported by a PMA Division
Fellowship at Caltech and NSF Grant AST-9417371.
WK is supported by the Deutscher Akademischer Austauschdienst
(DAAD). PL is supported by the Binary Black Hole Grand Challenge
Alliance, NSF PHY/ASC 9318152 (ARPA supplemented) and by NSF grants
PHY 96-01413, 93-57219 (NYI).

%%%%%%%%%%%%%%%%%%%%%%%%%%%%%%%%%%%%%%%%%%%%%%%%%%%%%%%%%%%%%
%  BIBLIOGRAPHY
%%%%%%%%%%%%%%%%%%%%%%%%%%%%%%%%%%%%%%%%%%%%%%%%%%%%%%%%%%%%%

%%%%%%%%%%%%%%%%%%%%%%%%%%%%%%%%%%%%%%%%%%%%%%%%%%%%%%%%%%%%%%%%%%%%%%
%  END OF DOCUMENT
%%%%%%%%%%%%%%%%%%%%%%%%%%%%%%%%%%%%%%%%%%%%%%%%%%%%%%%%%%%%%%%%%%%%%%

\end{document}